# A hot compact dust disk around a massive young stellar object

Stefan Kraus[1], Karl-Heinz Hofmann[2], Karl M. Menten[2], Dieter Schertl[2], Gerd Weigelt[2], Friedrich Wyrowski[2], Anthony Meilland[2,3], Karine Perraut[4], Romain Petrov[3], Sylvie Robbe-Dubois[3], Peter Schilke[5] & Leonardo Testi[6]

[1]Department of Astronomy, University of Michigan, 500 Church Street, Ann Arbor, Michigan 48103, USA. [2]Max-Planck-Institut für Radioastronomie, Auf dem Hügel 69, 53121 Bonn, Germany. [3]Laboratoire Hippolyte Fizeau, UMR 6525, Université de Nice Sophia-Antipolis CNRS, Observatoire de la Côte d'Azur, Parc Valrose, 06108 Nice cedex 2, France. [4]Laboratoire d'Astrophysique de Grenoble, UMR 5571, Université Joseph Fourier CNRS, BP 53, 38041 Grenoble cedex 9, France. [5]I. Physikalisches Institut, Universität zu Köln, Zülpicher Strasse 77, 50937 Cologne, Germany. [6]INAF-Osservatorio Astrofisico di Arcetri, Largo Fermi 5, 50125 Firenze, Italy.

**Circumstellar disks are an essential ingredient of the formation[1] of low-mass stars. It is unclear, however, whether the accretion-disk paradigm can also account for the formation of stars more massive than about 10 solar masses[2], in which strong radiation pressure might halt mass infall[3,4]. Massive stars may form by stellar merging[5], although more recent theoretical investigations suggest that the radiative-pressure limit may be overcome by considering more complex, non-spherical infall geometries[6,7]. Clear observational evidence, such as the detection of compact dusty disks[8] around massive young stellar objects, is needed to identify unambiguously the formation mode of the most massive stars. Here we report near-infrared interferometric observations that spatially resolve the astronomical-unit-scale distribution of hot material around a high-mass (~20 solar masses) young stellar object. The image shows an elongated structure with a size of ~13 × 19 astronomical units, consistent with a disk seen at an inclination angle of ~45°. Using geometric and detailed physical models, we found a radial temperature gradient in the disk, with a dust-free region less than 9.5 astronomical units from the star, qualitatively and quantitatively similar to the disks observed in low-mass star formation. Perpendicular to the disk plane we observed a molecular outflow and two bow shocks, indicating that a bipolar outflow emanates from the inner regions of the system.**

We investigated the massive young stellar object IRAS 13481-6124, which harbours a central object of about 20 solar masses ($20M_{\odot}$ (ref. 9) embedded in a cloud with a total gas mass of ~1,470 $M_{\odot}$ (ref. 10). Our imaging observations covered wavelengths from 2 to 870 μm and spatial scales from several parsecs to astronomical units (AU). In archival Spitzer Space Telescope images[11] (Fig. 1a), we discovered two faint bow shocks, which are oriented along position angle PA = 31 ± 6° and separated by 7', corresponding to a physical scale of ~6.5 pc at a distance of 3,500 pc (ref. 12). We also observed the source with the Atacama Pathfinder Experiment (APEX) 12-m submillimetre telescope, located at the Llano de Chajnantor Observatory in Chile. Using observations in emission lines of molecular gas, we detected a bipolar outflow, which is oriented along the same axis (PA = 26 ± 9°, Fig. 1b) as the bow shocks. The detected outflow signatures have a relatively narrow bow-shock opening angle of ~6° and show similarities with the bipolar collimated jets observed in low-mass star formation. Outflows from young stars are powered by accretion[13] and require a magnetized, compact accretion disk to collimate the ejected material[14], so the detected outflow provides some indirect evidence that there is a disk around IRAS 13481-6124.

In order to directly detect this disk and to characterize its inner structure, we used the Very Large Telescope Interferometer (VLTI) of the European Southern Observatory (ESO), located on Cerro Paranal in Chile. Using the near-infrared beam combination instrument AMBER[15], we combined the light of three of the VLTI 1.8-m telescopes and measured the interferometric observables (visibilities and closure phases) in 17 spectral channels in the K-band (wavelength, $\lambda$ = 1.95–2.55 μm). We obtained the VLTI observations using three different three-telescope array configurations and complemented them with speckle-interferometric observations from the New Technology Telescope, providing precise spatial information for baseline lengths of less than 3.5 m.

Using our extensive VLTI + speckle data set, we reconstructed[16] a model-independent interferometric image from the measured visibility amplitudes and closure phases. This technique, which is routinely applied in radio interferometry, has recently been demonstrated with the VLTI[17]. We applied it in our study to image the circumstellar environment around a young star. With baseline lengths $B$ up to 85 m, our image provides an angular resolution of $\lambda/2B = 2.4$ milli-arcsecond (mas) or ~8.4 AU, which is at least one order of magnitude higher than conventional infrared imaging techniques at 10-m-class telescopes; the gain compared to state-of-the-art submillimetre disk studies is about two orders of magnitude. The reconstructed image (Fig. 1c) clearly resolves the inner environment around IRAS 13481-6124 and reveals a compact elongated structure with a size of 5 mas × 8 mas. The disk orientation (PA = 120°) is perpendicular to the determined outflow axis (26 ± 9° and 31 ± 6°), suggesting that the AU-scale disk resolved by VLTI/AMBER is indeed the driving engine of the detected collimated outflow.

To characterize further the detected elongated structure, we fitted geometric and detailed physical models to our data, allowing us to deduce object information smaller than the diffraction-limited resolution of our reconstructed image. Our measurements show that the visibility function drops rapidly at baseline lengths of less than 3 m, but then stays nearly constant up to baseline lengths of about 12 m, followed by a uniform, nearly linear decline at baseline lengths up to 85 m (Supplementary Fig. 10), providing clear evidence that there are at least two spatial components. We used a Gaussian model to estimate the characteristic size (full-width at half-maximum diameter) of the two components and found that the first, extended component has a size of about 108 mas and contributes approximately 15% of the total flux. On the basis of the discovery of scattered-light envelopes around other massive stars[18], we suggest that this extended centrosymmetric component is the scattered-light contribution of a natal envelope. The second, compact component has a size of 5.4 mas, accounts for 85% of the total K-band flux and is significantly elongated along PA = 114°. Furthermore, we found that the size of the compact component increases significantly towards longer wavelengths, as shown by the measured equivalent widths in Fig. 2a. This effect, which was found in many low- and intermediate-mass young stellar object disks[19,20], probably indicates a temperature gradient in the circumstellar material, in which hotter material (radiating more effectively at shorter wavelengths) is located closer to the star.

Motivated by these indications for an internal temperature structure, we fitted analytical disk models with a radial-temperature power-law $T(r) = T_{in} (r/r_{in})^{-q}$, where $T(r)$ is the temperature at radius $r$, $T_{in}$ is the inner-disk temperature, $r_{in}$ is the inner-disk radius and $q$ is the temperature power-law index, to our data. Assuming reasonable values for $T_{in}$ of 1,500–2,000 K, we found that this model could reproduce our interferometric data with an $r_{in}$ of 3 mas (9.5 AU) and a $q$ of 0.4, which is consistent with the theoretical temperature gradient[21] of flared irradiated disks ($q = 0.43$). This disk model provides a significantly better representation of our data than do other geometric models or disk models with emissions extending to the star (Fig. 2b and Supplementary Information Section 4). The derived inner-disk radius of 9.5 AU agrees with the expected location of dust sublimation in an irradiated circumstellar disk: 6.2–10.9 AU, assuming grey dust and dust-sublimation temperatures between 1,500 and 2,000 K. Therefore, IRAS 13481-6124 follows the well established size–luminosity relation[22] for low- to intermediate-mass young stellar objects, suggesting that the near-infrared emission mainly traces material at the dust sublimation radius, similarly to the disks around T Tauri and Herbig Ae/Be stars.

In addition to the rather general arguments presented above, we applied detailed physical models using multi-dimensional continuum radiative-transfer simulations. Encompassing the central star, disk and envelope, these models can provide a much more comprehensive and consistent picture of the IRAS 13481-6124 system than purely geometric models. Because these radiative-transfer models depend on many free parameters, we used a precomputed model grid[23] and an associated spectral energy distribution (SED)-fitting tool[24] to find a good initial set of model parameters. Using an adapted version of the radiative-transfer modelling code by Whitney et al.[25] we further refined the model parameters and found a model that could reproduce the SED and the VLTI + speckle observations simultaneously, yielding the model image shown in Fig. 1d (see Supplementary Figs 13 and 14 for visibilities, closure phases and the SED, and Supplementary Table 4 for the complete set of model parameters). Our global best-fit model parameters suggest a distance of 3,260 pc and include a central star with a mass of $18 M_\odot$, a massive ($1,000 M_\odot$) circumstellar envelope with a bipolar curve-shaped outflow cavity and a compact (~130 AU) flared disk with a mass similar to that of the central star ($20 \pm 8 M_\odot$). Many model parameters, such as those concerning the envelope geometry and the disk mass, may still suffer from the parameter ambiguities that are inherent to SED-fitting results[24]. Other parameters, such as the inner-disk radius and the disk orientation, are constrained by spatially resolved observations. The resolved disk lacks asymmetry, as is evident from the measured small closure phases and the

Page 4 of 11

symmetrical reconstructed image. For our radiative-transfer models, this low degree of asymmetry puts direct constraints on the vertical disk structure near the dust-sublimation region and suggests that the inner dust rim is smooth. As with findings for intermediate-mass stars, this probably indicates that the inner dust rim around IRAS 13481-6124 is curved (see also Supplementary Information Sections 3 and 5), although further theoretical and observational work is needed to characterize fully the detailed inner-rim geometry. In particular, future observations using mid-infrared and submillimetre interferometers could resolve the inner-disk structure over a very wide wavelength range, putting more constraints on the radial density structure and the more extended disk regions.

The measured low degree of asymmetry is also interesting in the context of numerical simulations of self-gravitating disks, which predict the formation of large-scale spiral density waves[8,26] and a highly non-axisymmetric disk structure. Given that these asymmetries are not detected by our VLTI phase-closure observations, we conclude that these processes are not dominant in the probed inner regions of the IRAS 13481-6124 disk, or are not sufficiently resolved. Our speckle and VLTI data also rule out the presence of nearby stellar companions (down to flux ratios of ~1:40 and separations exceeding 10 mas), contrasting with the prediction of compact stellar clusters in competitive accretion and stellar merger models[27]. Our observations suggest that IRAS 13481-6124 has already gone through the main accretion phase in its evolution, and that the disk is heated mostly by stellar irradiation. The derived disk structure, which closely resembles the irradiated dust disks around Herbig Ae/Be stars but is distinctly different from the actively accreting disks observed around FU Orionis stars, for instance, supports this conclusion. According to our radiative-transfer model and the underlying stellar evolutionary tracks[28], the system has an evolutionary age of about 60,000 years, roughly matching the expected age for the start of disk evaporation, 100,000 years (ref. 29). Therefore, we are probably observing the system in a short-lived phase, in which the strong stellar winds, radiation pressure and photoevaporation have already stopped the mass infall, and are just starting to dissipate the circumstellar disk.

**Supplementary Information** is linked to the online version of the paper at www.nature.com/nature.

**Acknowledgements** This work was done in part under contract with the California Institute of Technology (Caltech), funded by NASA through the Sagan Fellowship Program (S.K. is a Sagan Fellow). We thank the ESO Paranal staff for support and their efforts in improving the VLTI. This paper is based on observations made with ESO telescopes at the La Silla Paranal Observatory and archival data obtained with the Spitzer Space Telescope, operated by the Jet Propulsion Laboratory, Caltech, under a contract with NASA. We also used data acquired with APEX, a collaboration between the Max-Planck-Institut für Radioastronomie, ESO, and the Onsala Space Observatory.

**Author Contributions** S.K. worked on the AMBER data reduction and data interpretation, model fitting and image reconstruction, made some of the observations and wrote the telescope proposals and the initial paper manuscript. K.-H.H. worked on the image reconstruction. D.S. worked on the speckle data reduction. G.W. made some of the observations. F.W. worked on the APEX data reduction and data interpretation. K.M.M. and P.S. were co-authors on the telescope proposal. A.M., K.P., R.P., S.R.-D. and L.T. represent the AMBER instrument consortium, which provided some observation time. All authors commented on the paper.

**Author Information** Reprints and permissions information is available at www.nature.com/reprints. The authors declare no competing financial interests. Readers are welcome to comment on the online version of this article at www.nature.com/nature. Correspondence and requests for materials should be addressed to S.K. (stefankr@umich.edu).

**Figure 1 Zoom in on IRAS 13481-6124, covering structures over more than five orders of magnitude. a**, In Spitzer/Infrared Array Camera (IRAC) images, we detect two bow-shock structures, indicating a collimated bipolar outflow. The bows (insets) are separated by 7′ and appear as excess emissions in the IRAC 4.5-μm band (green), probably tracing shocked molecular hydrogen gas[30]. The colours in the composite image correspond to wavelength-bands around 3.6 μm (blue), 4.5 μm (green), and 8 μm (red). **b**, The outflow is also detected in molecular line emissions using the APEX/Swedish Heterodyne Facility Instrument on scales of a few 10,000 AU, with the approaching (blueshifted) lobe southwest of IRAS 13481-6124 (solid contours show a blueshifted velocity of −10 km s$^{-1}$ and dashed contours a redshifted velocity of +5 km s$^{-1}$). We used the $^{12}$CO (3–2) emission line at 867 μm as the outflow tracer line. **c**, VLTI/AMBER aperture-synthesis imaging reveals an elongated structure perpendicular to the outflow direction (contours decrease from peak intensity by factors of √2). The structure has a size of 5 mas × 8 mas, as measured at 10% of the peak flux intensity, and contributes about 80% of the total flux in the image; the remaining flux elements are spread uniformly over the image and correspond to the extended component detected in our model-fits. The image was reconstructed from an extensive data set of 33 independent VLTI/AMBER three-telescope observations obtained around a wavelength of 2.2 μm and using three different three-telescope array configurations. **d**, The best-fit radiative-transfer model image as constrained by the measured wavelength-dependent visibilities and closure-phases and the SED from near-infrared to millimetre wavelengths (see also Supplementary Figs 13 and 14). The image was computed for a wavelength of 2.2 μm and brighter colours indicate higher flux. This modelling allows us to determine the intensity profile on scales smaller than the diffraction-limited resolution, and strongly suggests that the dust disk is truncated at a radius of 6.2 ± 1.2 AU.

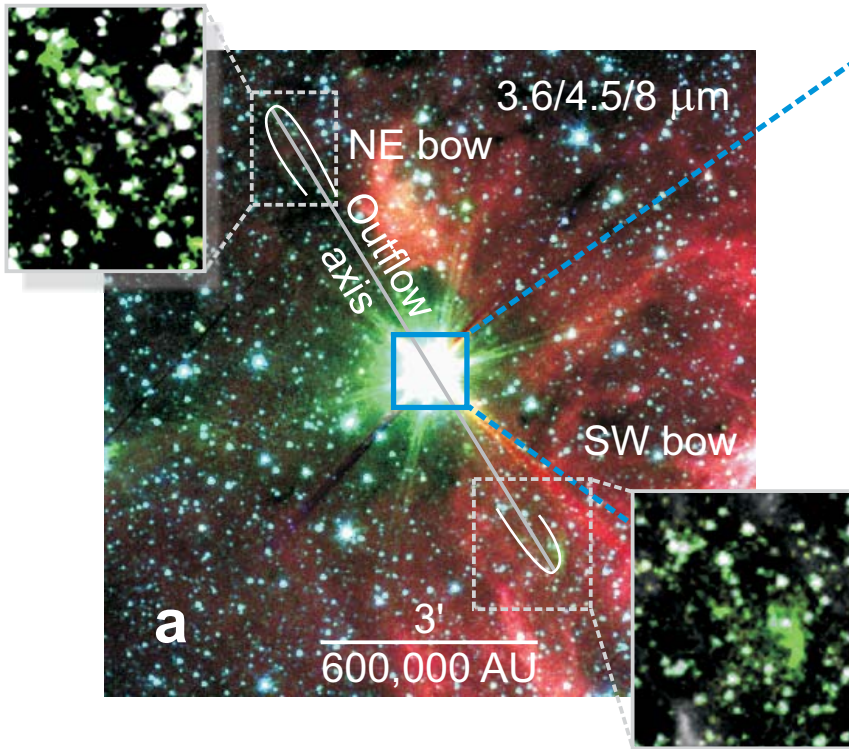
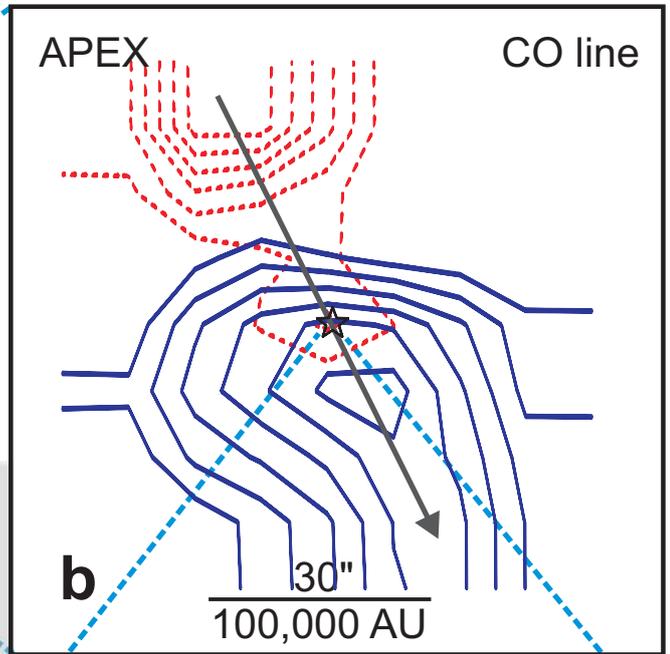
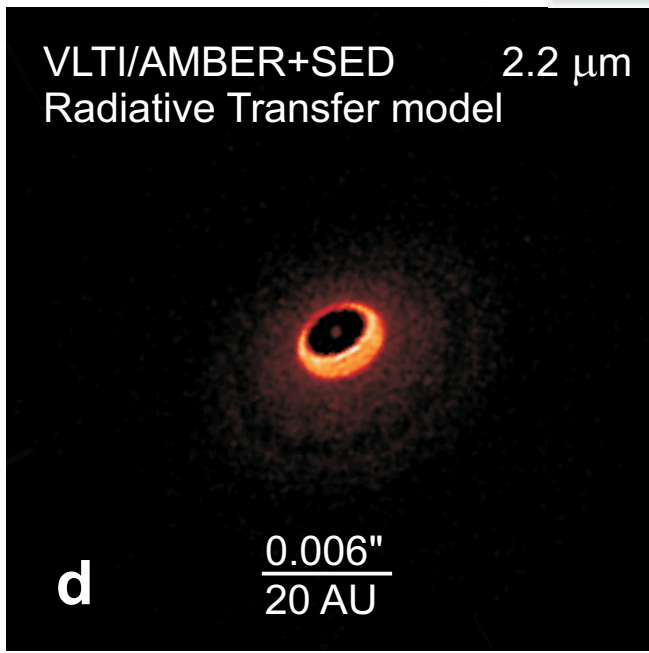
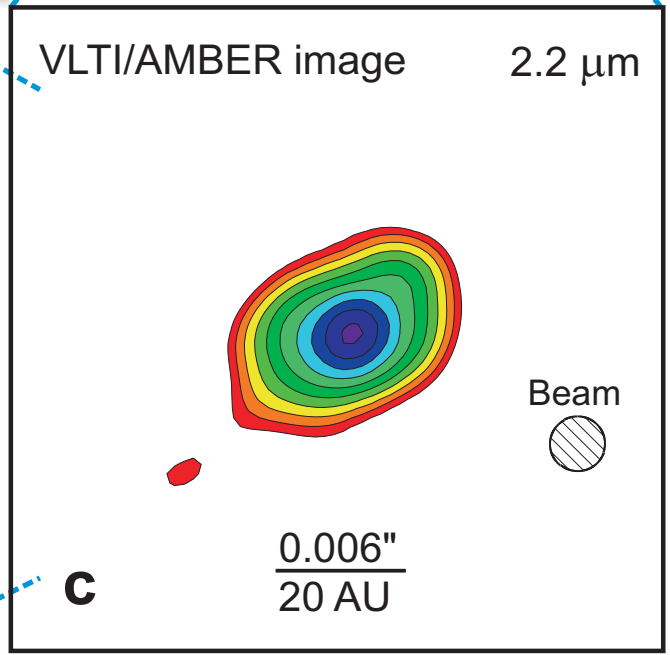

**Figure 2 Elongation of the compact emission component, as determined with a Gaussian and a temperature-gradient disk model. a**, Using a Gaussian model, we determined the object size for different position-angle bins (each covering 10°) and wavelength channels centred on 2.1 μm (blue), 2.3 μm (green) and 2.5 μm (red), revealing strong object elongation and indications of a temperature gradient. The error bars give the standard deviation. **b**, Using a temperature power-law disk-model $T(r) = 2{,}000\,\text{K}\,(r/r_{\text{in}})^{-q}$, we determined the temperature power-law index $q$ to be ~0.4 and found an ellipsoidal source geometry, corresponding to a disk seen under an inclination angle of ~45°. The truncated dust-disk model yields a significantly better representation of the measured wavelength-dependent visibilities ($\chi^2$/d.f. = 1.1) than standard geometric models (for example, Gaussians, $\chi^2$/d.f. = 2.5, or disks of uniform brightness, $\chi^2$/d.f. = 4.9; see Supplementary Information Section 4).

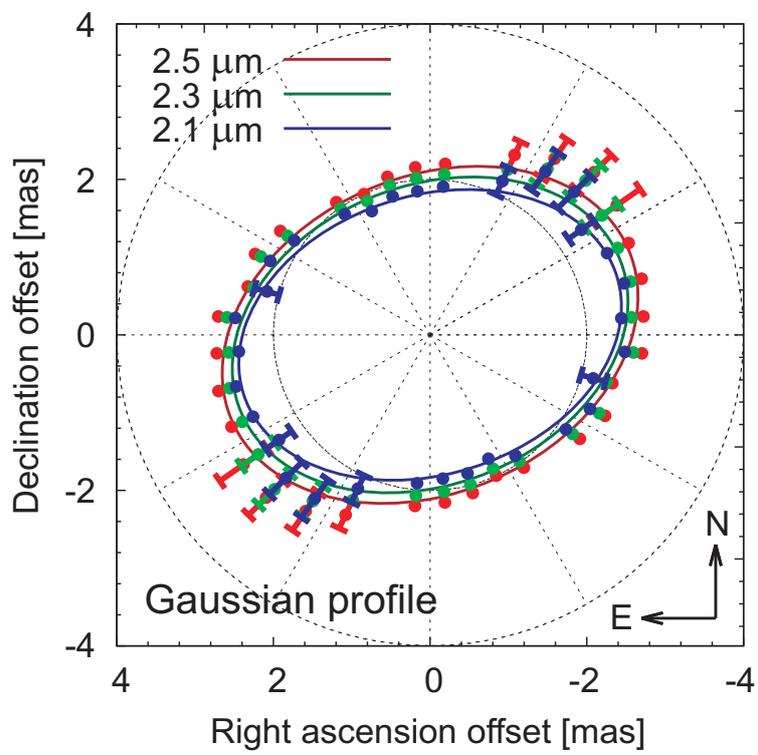 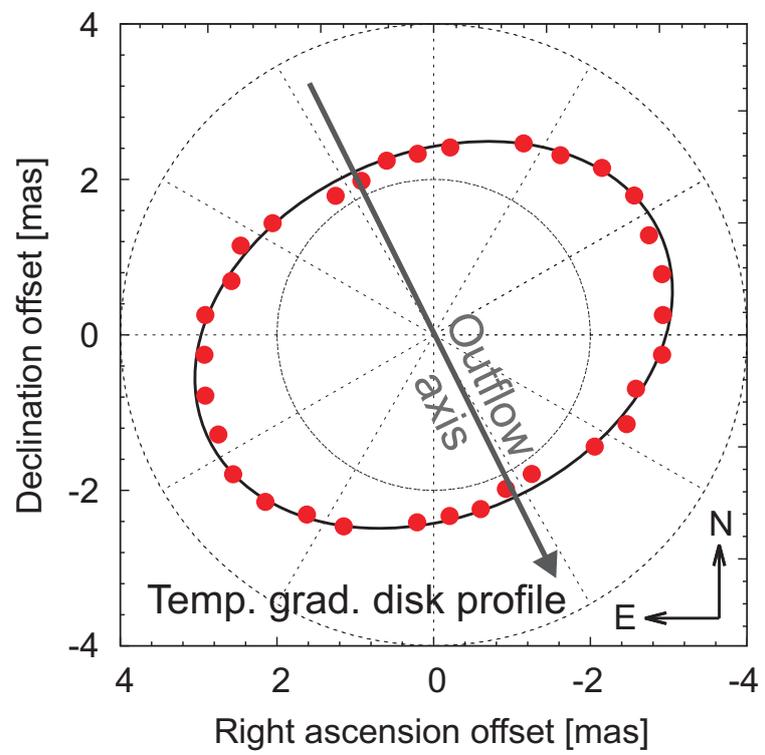